\documentclass[a4paper,11pt]{article}
\usepackage{pos}
\usepackage{graphicx}
\usepackage{subfig}
\usepackage{mathtools}
\DeclarePairedDelimiter\ceil{\lceil}{\rceil}

\newcounter{inlineenum}
\renewcommand{\theinlineenum}{\alph{inlineenum}}
\newenvironment{inlineenum}
  {\unskip\ignorespaces\setcounter{inlineenum}{0}%
   \renewcommand{\item}{\refstepcounter{inlineenum}{\textit{\theinlineenum})~}}}
  {\ignorespacesafterend}

\newcounter{daggerfootnote}
\newcommand*{\daggerfootnote}[1]{%
    \setcounter{daggerfootnote}{\value{footnote}}%
    \renewcommand*{\thefootnote}{\fnsymbol{footnote}}%
    \footnote[2]{#1}%
    \setcounter{footnote}{\value{daggerfootnote}}%
    \renewcommand*{\thefootnote}{\arabic{footnote}}%
    }

\title{Cosmic-Ray Composition analysis at IceCube using Graph Neural Networks}
\manuallySeparateAuthors
\author*[a]{Paras Koundal }
\author[]{ for the IceCube collaboration \protect\daggerfootnote{Full Author List : \url{https://icecube.wisc.edu/collaboration/authors/}}}

\affiliation[a]{Institute for Astroparticle Physics, Karlsruhe Institute of Technology,\\
   Hermann-von-Helmholtz-Platz 1, 76344 Eggenstein-Leopoldshafen, Karlsruhe, Germany}

\emailAdd{paras.koundal@kit.edu}
\abstract{The IceCube Neutrino Observatory is a multi-component detector embedded deep within the South-Pole Ice. This proceeding will discuss an analysis from an integrated operation of IceCube and its surface array, IceTop, to estimate cosmic-ray composition. The work will describe a novel graph neural network based approach for estimating the mass of primary cosmic rays, that takes advantage of signal-footprint information and reconstructed cosmic-ray air shower parameters. In addition, the work will also introduce new composition-sensitive parameters for improving the estimation of cosmic-ray composition, with the potential of improving our understanding of the high-energy muon content in cosmic-ray air showers.}

\FullConference{%
  *** 27th European Cosmic Ray Symposium - ECRS ***\\
  *** 25-29 July 2022 ***\\
  *** Nijmegen, the Netherlands ***
}

\begin{document}
\maketitle

%==============================%
\section{Introduction}
\begin{figure}[!h]
\centering
\includegraphics[width=0.23\textheight,origin=c]{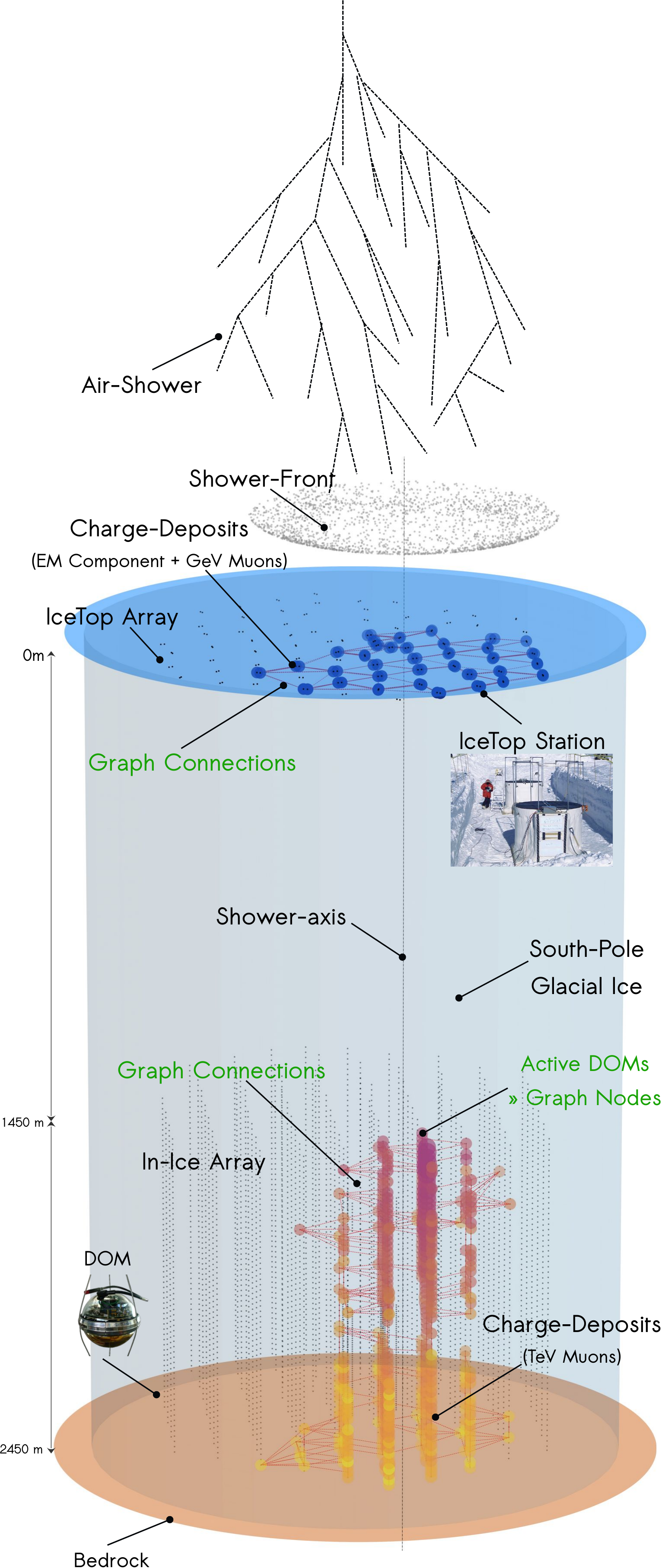}
\caption{An illustration of an example EAS incident on IceTop. The dark-blue(/colored) circles represent the station(/DOM) hits after spatio-temporal cleaning around the reconstructed shower-axis. Only the hits are interpreted as nodes of a graph and the graph-connections  are based on a predefined/adaptive algorithm.}
\label{fig:IceCube_Shower_GNN}
\end{figure}

IceCube is a cubic-kilometer astroparticle detector at the geographic South Pole. Over 5000 digital optical modules (DOMs) are deployed on 86 strings in glacial ice at depths ranging from 1450 m to 2450 m \cite{aartsen2017icecube}. It detects particles from astrophysical sources, and tries to understand the dynamics of the sources through their cosmic ambassadors, allowing analysis in the areas of cosmic rays, neutrino physics, and other research areas. The in-ice IceCube (IC) array is accompanied by the surface component, called IceTop (IT) \cite{abbasi2013icetop}. In addition to this IceCube also has a DeepCore (DC) array with DOM density roughly five times higher than that of the standard IC array, located around the center of the IC array at depths below 2100 m \cite{abbasi2012design}. This makes it a unique three-dimensional multimessenger detector. IT is predominantly used as an extensive air-shower (EAS) detector, in addition to serving as a veto for neutrino detection. In an integrated approach, IT (primarily detects the electromagnetic component + GeV muons) and IC+DC (primarily detects TeV muons) can be used to reconstruct the direction, energy, and mass of penetrating particles of the incident cosmic ray (CR). An illustration of an example air shower incident on IT with also a footprint in IC is shown in \autoref{fig:IceCube_Shower_GNN}.\par  

IceCube has already demonstrated the ability to improve our understanding of the energy and mass spectrum in the transition region from galactic to extragalactic cosmic rays \cite{aartsen2019cosmic}.
This work extends ongoing analyses \cite{koundal2021study, koundal2022composition}, where the cleaned signal footprint is used. This has been possible because of ongoing advancements in the field of deep-learning. For the estimation of primary mass a graph neural network (GNN) based approach has been used. The Monte-Carlo (MC) simulations used for the analysis and training were generated using the CORSIKA air-shower generator \cite{heck1998corsika}. The datasets for each primary type (p, He, O and Fe) were simulated in the energy range 5.0 $\leq$ log$_{10}$(E/GeV) $\leq$ 8.0, using FLUKA \cite{battistoni2007fluka} as the low-energy hadronic interaction model and SIBYLL 2.1 \cite{ahn2009cosmic} as the high-energy interaction model. For this investigation, we only consider coincident IT+IC events utilizing quality cuts \cite{aartsen2019cosmic} that guarantee coincidence, with successful direction reconstruction and shower-core containment within the IT array. This work will also introduce two new composition-sensitive parameters.

%==============================%
\section{New Composition-Sensitive Parameters} \label{sec:NewParams}
IceCube has performed multiple analyses, where several shower-features using simulations were  probed \cite{koundal2021study, koundal2022composition} and validated on real data \cite{aartsen2019cosmic}. $ \mathbf{log_{10}(S_{125} / VEM)} $, where $S_{125}$ is the signal expectation at a perpendicular distance of 125 m from the shower axis at IT, was found to be a good energy-proxy \cite{klepser2008reconstruction}. The signal is in units of Vertical Equivalent Muon (VEM), which is the signal that
would be deposited by a single muon vertically traveling through an entire IT tank. The charge deposits at IT %for reconstructing energy and composition proxies, they 
are also used to reconstruct directional information like zenith and azimuth \cite{aartsen2019cosmic}. For mass-discrimination we had developed multiple useful parameters \cite{aartsen2019cosmic, pandya2019search, medina2021reconstruction, koundal2021study,koundal2022composition}. Among those, the ones used in this study are
\begin{inlineenum}
\item \( \mathbf{log_{10}(dE/dX_{1500~m})} \) = fit-value of IC energy-loss profile at a slant-depth of 1500 m (details in \cite{aartsen2019cosmic})
\item \textbf{total stochastic energy} = total IC energy of high-energy local-stochastic deposits in an event (details in \cite{koundal2021study}).
\end{inlineenum} 
These parameters primarily probe energy-deposits by TeV muons in the detector.
In the following we discuss ongoing efforts to develop newer composition sensitive parameters.

\begin{figure}[!h]
\subfloat{\includegraphics[width=.50\textwidth]{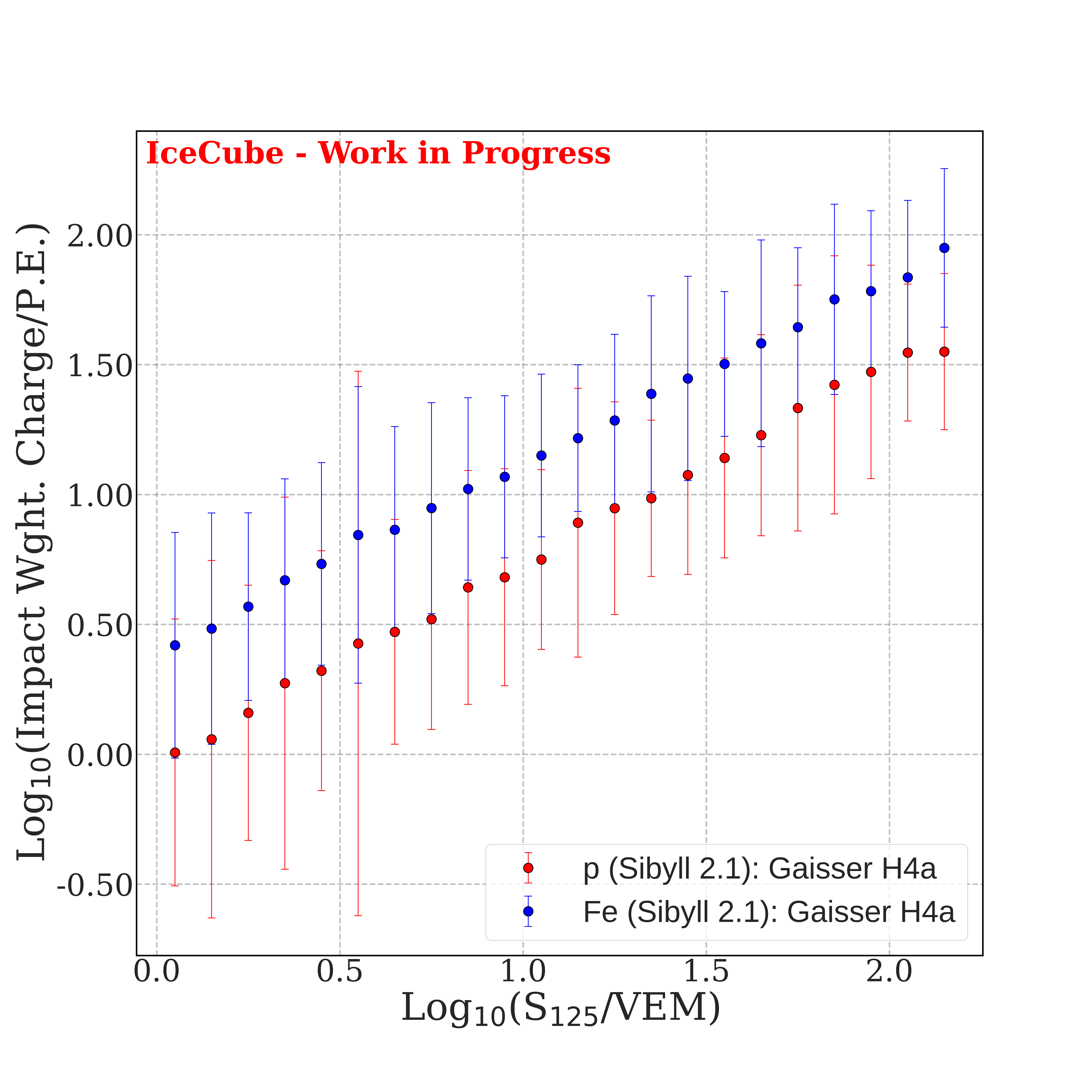}}
\subfloat{\includegraphics[width=.50\textwidth]{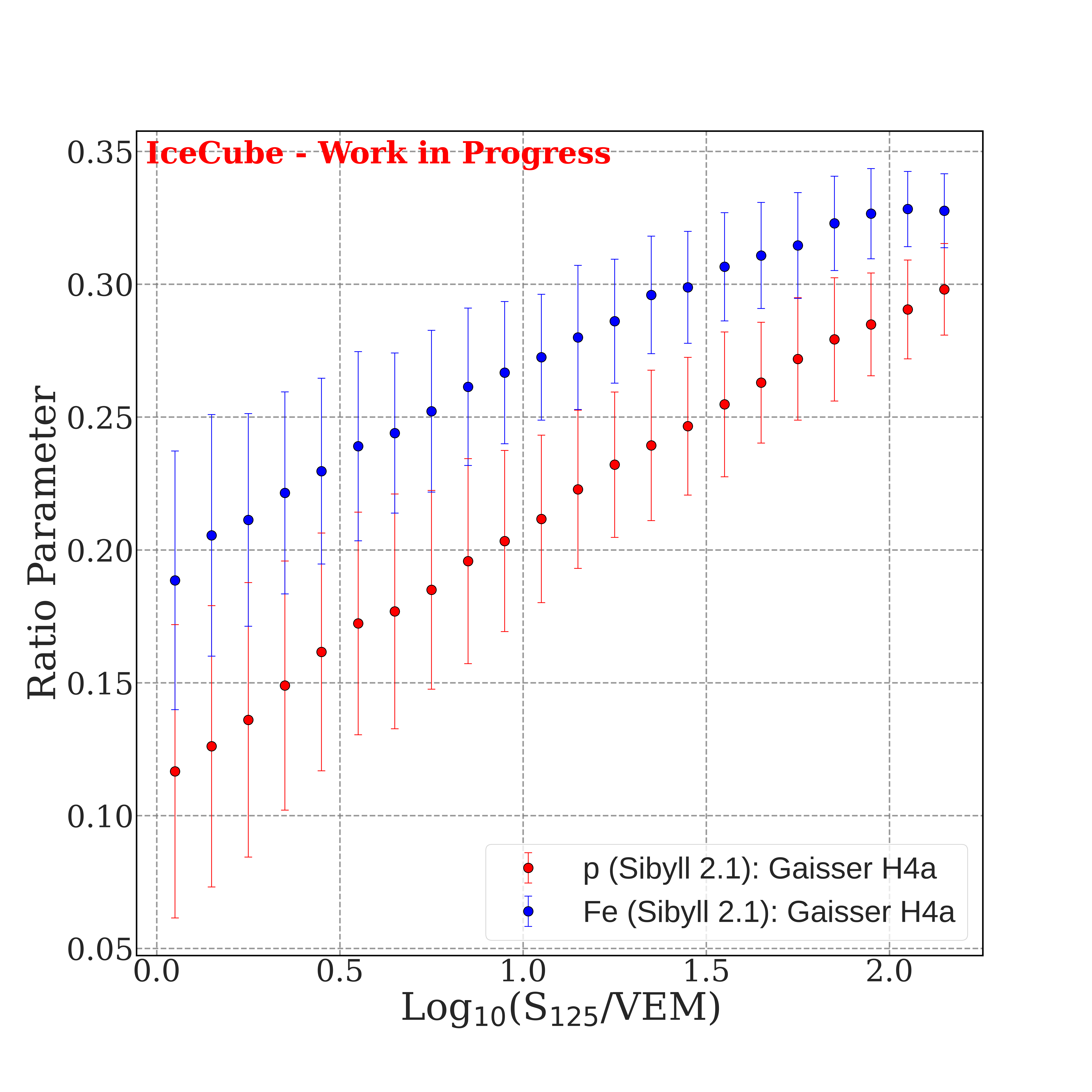}}
\caption{ Composition sensitivity for new IceCube \& IceTop variables (using SIBYLL 2.1 \cite{ahn2009cosmic} and weighted to the H4a all-particle spectrum \cite{gaisser2012spectrum}): \textit{Left:} Impact-Weighted Charge (last percentile bin), \textit{Right:} Ratio Parameter; as a function of shower size, $S_{125}$, for proton and iron primaries.}
\label{fig:new_mass_variables}
\end{figure}
%==============================%

\subsection{Impact-Weighted Average Charge (IAC)}
The information about the lateral-spread of in-ice muons was ignored in the previous studies. However, from our understanding of air-shower physics, at the same primary-energy Fe-induced air-showers are expected to have wider muon bundles than those initiated by p. This is anticipated because of aggregated behaviour of two observations from air-shower physics, namely at same energy:
\begin{inlineenum}
\item Fe initiated showers interact earlier in the atmosphere than p.
\item Fe initiated showers have larger muon-multiplicity with lower average muon energy.
\end{inlineenum}
For Fe showers, these phenomena allow for lower energy-muons with larger transversal momenta to be situated further away from the shower axis.\par
To test the hypothesis for IceCube, the IC spatio-temporal cleaned signal-footprint was divided along the slant-length into 5 percentile-bins (based on the number of DOM hits). The bins were divided based on percentiles (rather than depth) to ensure that we compare showers in the same zenith-bin (of the same energy and primary type) at approximately the same stage of shower development (independent of the point of first-interaction). This also prevents sparsity (and hence over-reliance on few DOMs) for the following parameter calculation. In each percentile bin the Impact-Weighted Average Charge (IAC) is calculated, and represented as:
\begin{equation}
    IAC_{Percentile~Bin} = \frac{\sum_{i} C_{i}r_{i}}{\sum_{i} r_{i}}
\end{equation}\par
where the summation runs over all the spatio-temporal cleaned hit-DOMs in a given percentile bin and $C_{i}$ represents the charge at DOM $i$ located at a perpendicular distance $r_{i}$ (impact) from the reconstructed shower-axis. It was found that the last percentile bin shows the maximal composition sensitivity. The results for p-Fe separation, for the last percentile bin, is shown in \autoref{fig:new_mass_variables} (left). As can be seen, the parameter shows a promising discrimination power for CR composition analysis. Further studies to improve the discrimination power of the parameter are ongoing. A future analysis can also check the feasibility of using the parameter for non-coincident showers.\par
This study also explored the usage of Charge-Weighted Average Impact (CAI) as a mass discriminator. It also showed mass discrimination, though much lower than IAC. Hence, the plots are not shown here. For the GNN analysis (\autoref{sec:GNN}), both IAC and CAI in all percentile bins were used.

%==============================%
\subsection{Ratio Parameter} \label{subsec:RatioParam}
KASCADE-Grande  was an air shower experiment located at the Karlsruhe Institute of Technology (average atmospheric depth 1022 g/cm$^2$) and has demonstrated that the ratio of muon-number to total charged-particle number is a good composition-sensitive parameter \cite{arteaga2011study, apel2013kascade}. IceCube lacks the capability of directly counting the particle-number for different shower-components. In order to test the feasibility of particle-type ratios for mass-discrimination at IceCube, reconstructed proxy-parameters were used to approximate the muon-number (using \( \mathbf{log_{10}(dE/dX_{1500~m})} \)) and total charged-particle number (using \( \mathbf{log_{10}(S_{125})} \) ). The preliminary results from such a test are shown in \autoref{fig:new_mass_variables} (right). As can be seen from the figure, the parameter shows good discrimination power for CR composition analysis. This simple test also shows the potential benefits of obtaining particle-number estimates for composition analysis. Improvement in the mentioned parameter for IceCube  and similar tests for IceCube-Gen2 are ongoing.

%==============================%
\section{Cosmic-Ray Composition : Using Graph Neural Networks}\label{sec:GNN}
The IceCube Collaboration plans to enhance the present detector to the  next-generation instrument called IceCube-Gen2 \cite{aartsen2021icecube}. The detector will enhance our understanding of the multimessenger universe by combining observations from multiple cosmic messengers. In addition to this, the enhancements will also extend the energy-sensitivity range and sky coverage of current measurements. The science objectives will be met by incorporating four new components: an in-ice optical array, a low-energy core, a surface air shower array \cite{haungs2019scintillator}, and an extended radio detector array. %However, as can be seen from \autoref{fig:IceCubeGen2}, 
The enhancements will also shift the detector geometry to a more irregular one. The viability of utilizing GNNs for making composition analysis flexible to such detector upgrades has been demonstrated in prior studies \cite{koundal2021study, koundal2022graph, koundal2022composition}. This work extends that by aggregating older and newer composition-sensitive parameters, along with other relevant shower parameters. In addition to that, there have been technical changes in the GNN architecture, in comparison to \cite{koundal2021study}. The full network is trained as a regression-model with the logarithmic mass of CR primary i.e. ln(A) as the expected output. %The network is trained on NVIDIA Tesla V100.

%==============================%
\subsection{Network Architecture} \label{subsec:GNN_Arch}
\begin{figure}[!h]
\centering
\includegraphics[width=0.50\textheight,origin=c]{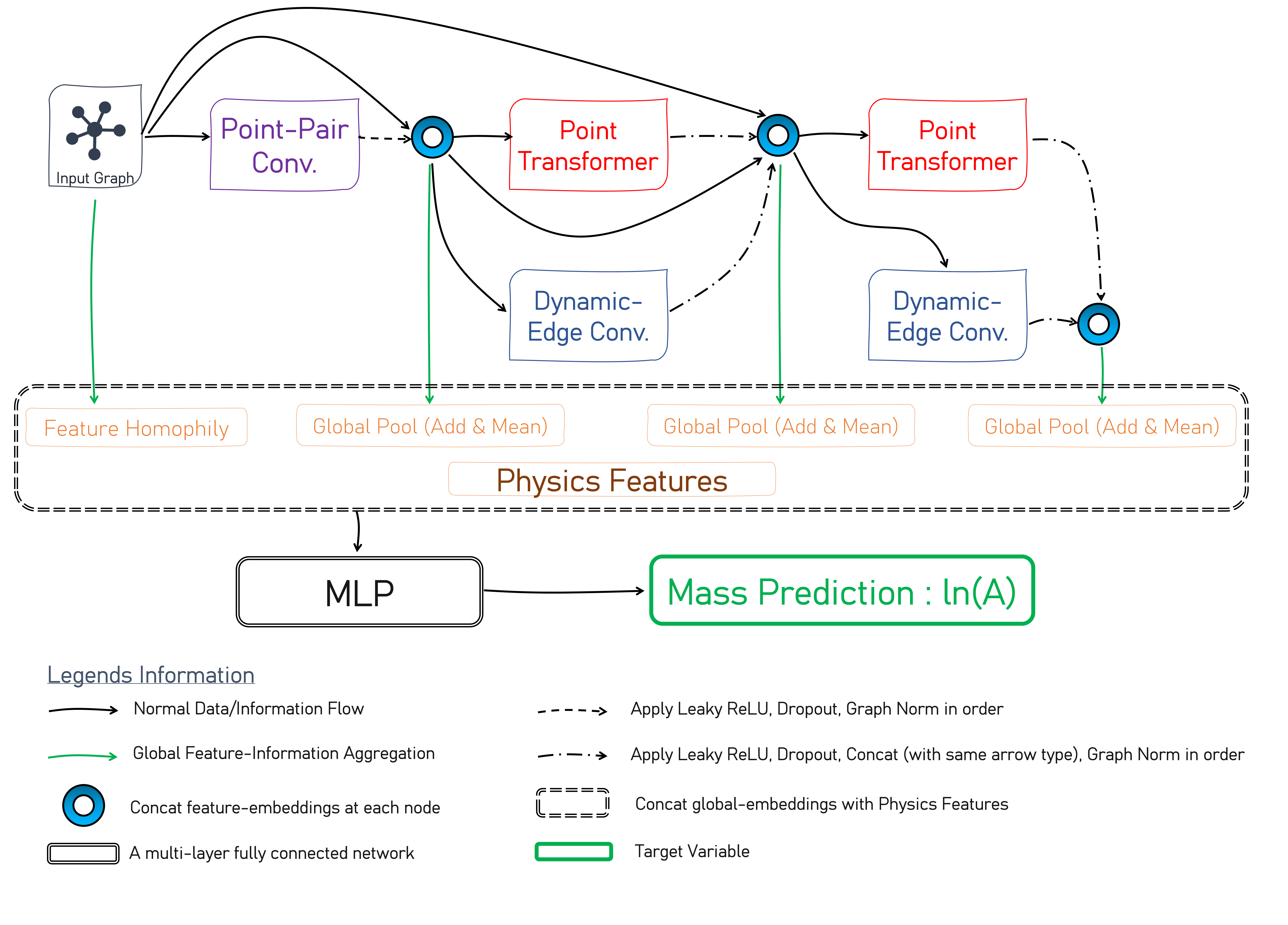}
\caption{Current GNN based architecture used for CR analysis at IceCube Observatory.}
\label{fig:GNN_arch}
\end{figure}
The current detector can easily be mapped to an orthogonal geometry, which was the reason for an extensive use of convolutional neural network (CNN) based approaches at IceCube. However, CNN for IceCube currently has a few minor problems like sparsity of DOM-hits (e.g. \autoref{fig:IceCube_Shower_GNN}) and treating DC \cite{abbasi2021convolutional, micallef2021using, yu2021direction} and IT \cite{IceCubeCollaborationSaffer2022_1000143047} separately. These problems will be exaggerated even more as we move to IceCube-Gen2. GNNs are used in order to develop a method which allays these problems and is flexible to detector upgrades. We overcome the sparsity problem by mapping only the active stations/DOMs as nodes of a graph. The problem of combining IT, IC and DC is automatically solved by using GNNs, since they allow us to learn on irregular point clouds of variable size. Mapping IT as a graph and using it together with IC+DC is also a major addition in contrast to the previous study \cite{koundal2021study}. Additionally, the physics features included
are \begin{inlineenum}
\item $\log_{10}(S_{125})$ 
\item $\log_{10}(dE/dX_{1500m})$
\item total stochastic energy 
\item IAC and CAI (all percentile bins)
\item Ratio Parameter
\item track length (In-Ice)
\item reconstructed Zenith and Azimuth
\end{inlineenum}.\par
To map the detector (IT+IC+DC) as a graph, the active stations/DOMs (hits remaining after signal cleaning) are mapped as nodes of the graph. Each node of the graph has associated attributes, which capture the spatial coordinates, the charge and timing information of the measured waveform for the hit DOMs. Defining node-neighbourhood i.e. edge-information at each node in the input-graph is another component essential for learning on graphs. Multiple methods were tested to define a neighbourhood. For the input-graph,\textbf{ weighted k-NN with inductive-bias} has proven to be the best option. In our context:
\begin{inlineenum}
    \item k-NN means connect each node to its k Nearest-Neighbors (in coordinate space) 
    \item weighted k-NN means weight the k-NN connections between two nodes by spatial separation between them.
    \item inductive-bias means the node connections indirectly capture the information of graph (or event) size and the primary-type (by scaling Ratio Parameter described in \autoref{subsec:RatioParam} - SRP).
\end{inlineenum}In summary, each node of an input-graph (or event) with N nodes has k neighbours, where:
\begin{equation} \label{eqn:NumConn}
    k =  1 + \ceil*{\frac{N}{1+ A \cdot exp[ SRP\cdot(B-B \cdot SRP \cdot N)]}}
\end{equation}\par 
where A and B are two positive-valued hyperparameters of the network. The detailed reasons and tests for choosing the form of \autoref{eqn:NumConn} will be published in a future publication.\par  
The steps described earlier give us an input graph (labelled as Input Graph in \autoref{fig:GNN_arch}), which can then be used as an input for the GNN architecture. Feature homophily gives a measure of how similar individual node-features of the connected-nodes are to each other \cite{zhu2020beyond}. The rest of the architecture is motivated from developments in powerful CNN architectures detailed in \cite{szegedy2015going, huang2017densely}, and are adapted for the use case of graphs. The convolution-types are based on studies with point clouds \cite{deng2018ppfnet, zhao2021point, wang2018dynamic}. The physics features are concatenated at the global pooling layer. This is finally fed into a multi-layer perceptron (MLP) with ln(A) as the target output.  

%==============================%
\subsection{Primary Type Discrimination using GNNs}\label{subsec:GNN_Results}
\begin{figure}[!h]
\centering
\includegraphics[width=0.55\textheight,origin=c,trim={0 0 10cm 0},clip]{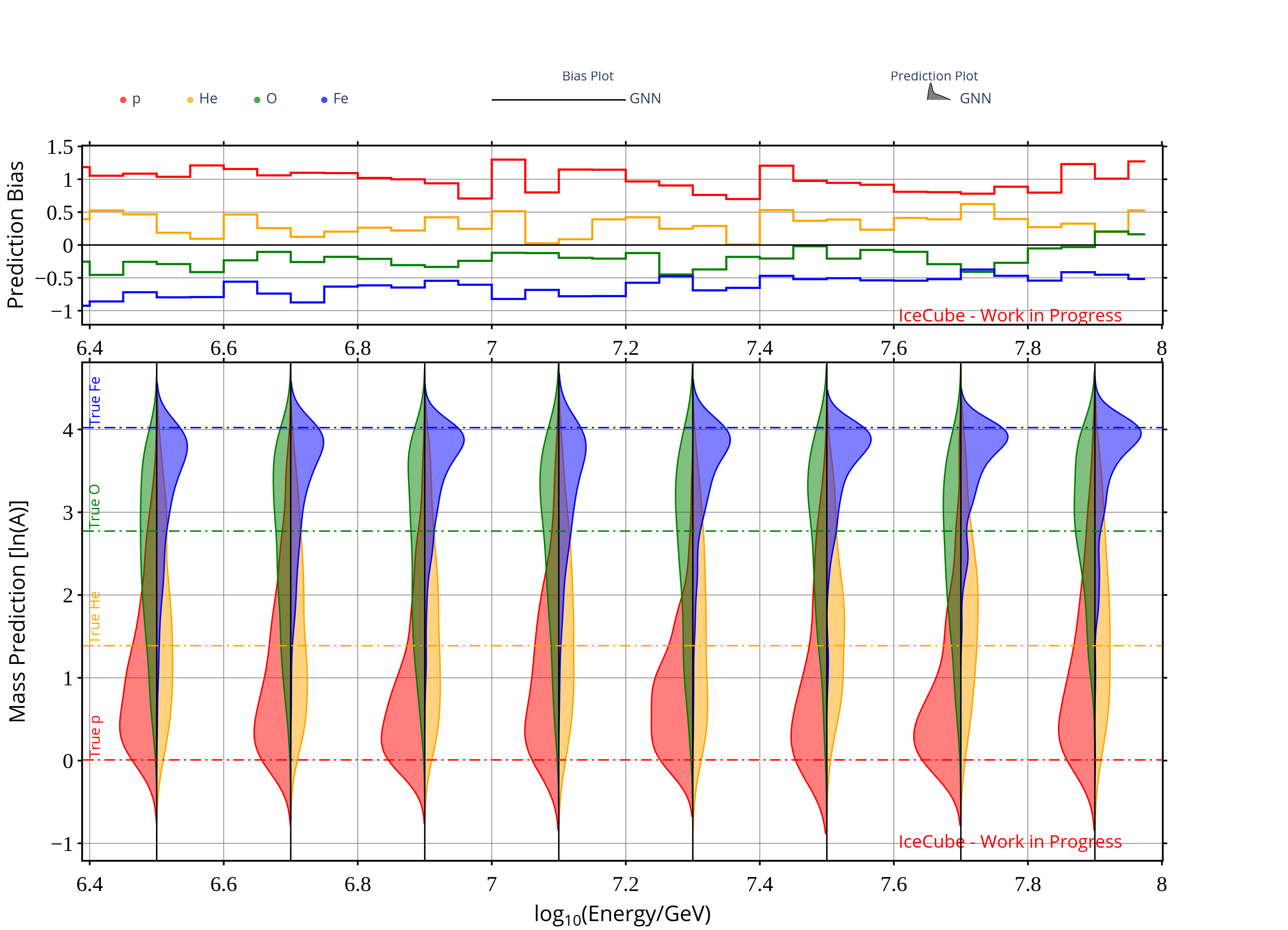}
\caption{Kernel density estimate (KDE) [Bottom] of mass prediction by GNN in true MC energy bins and Horizontal dashed-dot lines represent True Primary Mass [ln(A)]. Top plot shows bias in predictions.}
\label{fig:ElemSeparation}
\end{figure}
For a good mass discrimination, it is required that in all the energy bins the network-predictions for any two elements are maximally separated from each other. As can be seen from \autoref{fig:ElemSeparation}, the GNN based method shows good mass resolution and preciseness in prediction over almost all the energy bins (true MC energy shown here). p-Fe separation shows improvement with increase in energy. The overlap-area between normalized  p-Fe KDEs (\autoref{fig:ElemSeparation}) is a around 27\% for 6.4 $\leq$ log$_{10}$(E/GeV) $<$ 6.6 and improves to around 15\% in the last-bin. For intermediate primaries there is still scope for improvement. 

\section{Conclusion and Outlook}\label{sec:outlook}
We presented a graph neural network based approach for estimating the mass of primary cosmic rays in the energy range 5.0 $\leq$ log$_{10}$(E/GeV) $\leq$ 8.0. The new results are promising and allow to report individual spectra for elemental groups in the future. There are also ongoing efforts to increase the discrimination-power of the newer composition sensitive parameters. Since graphs can be used for irregular-shaped data this study can in the future also be adapted for the planned multi-component irregular-shaped IceCube-Gen2 and the related surface-enhacement \cite{haungs2019scintillator}.

%==============================% If you are reading this: Had to comment unfortunately because of page-limit.  
\begin{acknowledgments}
This work is done with the support and collaboration of Karlsruhe Institute of Technology (KIT), Germany, and IceCube Collaboration. The authors would also like to thank the organizers of the 27th European Cosmic Ray Symposium (ECRS 2022) conference for the opportunity of giving a talk. The authors acknowledge support by the High Performance and Cloud Computing Group at the Zentrum für Datenverarbeitung of the University of Tübingen, the state of Baden-Württemberg through bwHPC and the German Research Foundation (DFG) through grant no INST 37/935-1 FUGG.
\end{acknowledgments}
%==============================% Also, thanks for reading my proceeding.

\bibliographystyle{JHEP}
{\scriptsize
\bibliography{bib}}

\end{document}